\documentclass{ws-ijmpe}
\usepackage{epsfig}
\usepackage{subfigure}
\usepackage{amsmath}
\usepackage{amsfonts}
\usepackage{amssymb}
\usepackage{url}
\newcommand{\be}{\begin{equation}}
\newcommand{\ee}{\end{equation}}
\newcommand{\bea}{\begin{eqnarray}}
\newcommand{\eea}{\end{eqnarray}}

\begin{document}
\markboth{Takeshi Fukuyama, Tatsuru Kikuchi and Nobuchika Okada}
{The current problems of the minimal SO(10) GUT and their solutions}
%
\catchline{}{}{}{}{}

\title{The current problems of the minimal SO(10) GUT and their solutions
\footnote{Talk given at the International Workshop on Neutrino Masses
and Mixings Toward Unified Understanding of Quarks and Lepton Mass Matrices,
held at University of Shizuoka on December 17-19, 2006.}
}

\author{\footnotesize Takeshi Fukuyama}

\address{Department of Physics, Ritsumeikan University,
Kusatsu, Shiga, 525-8577, Japan\\
fukuyama@se.ritsumei.ac.jp}

\author{\footnotesize Tatsuru Kikuchi}

\address{Theory Division, KEK,
Oho 1-1, Tsukuba, Ibaraki, 305-0801, Japan\\
tatsuru@post.kek.jp}

\author{\footnotesize Nobuchika Okada}

\address{Theory Division, KEK,
Oho 1-1, Tsukuba, Ibaraki, 305-0801, Japan\\
Department of Particle and Nuclear Physics,
The Graduate University for Advanced Studies, \\
Oho 1-1, Tsukuba, Ibaraki, 305-0801, Japan\\
okadan@post.kek.jp}

\maketitle

\begin{history}
\received{(received date)}
\revised{(revised date)}
\end{history}
\begin{abstract}
This talk consists of two parts.
In part I we review how the minimal renormalizable supersymmetric 
SO(10) model, an SO(10) framework with only one ${\bf 10}$ and 
 one $\overline{\bf 126}$ Higgs multiplets 
 in the Yukawa sector, 
 is attractive because of its highly predictive power.
Indeed it not only gives a consistent predictions on neutrino oscillation
 data but also gives reasonable and interesting values for leptogenesis,
LFV, muon $g-2$, neutrinoless double beta decay etc.
However, this model suffers from problems 
 related to running of gauge couplings. 
The gauge coupling unification may be spoiled 
 due to the presence of Higgs multiplets 
 much lighter than the grand unification (GUT) scale. 
In addition, the gauge couplings blow up around the GUT scale 
 because of the presence of Higgs multiplets 
 of large representations. 
In part II we consider the minimal SO(10) model in the warped extra dimension 
 and show a possibility to solve these problems.

\end{abstract}
\section{Part I}
The successful gauge coupling unification 
 in the minimal supersymmetric standard model (MSSM), 
strongly supports the emergence of a supersymmetric (SUSY) GUT 
 around $M_{\rm GUT} \simeq 2 \times 10^{16}$ GeV. 
SO(10) is the smallest simple gauge group 
 under which the entire SM matter content of each generation 
 is unified into a single anomaly-free irreducible representation, 
${\bf 16}$ representation. 
This ${\bf 16}$ representation includes 
 right-handed neutrino and SO(10) GUT incorporates the see-saw 
mechanism [\refcite{see-saw}].  
 Among several models based on the gauge group SO(10), 
 the renormalizable minimal SO(10) model has been paid 
 a particular attention, where two Higgs multiplets 
 $\{{\bf 10} \oplus {\bf \overline{126}}\}$ 
 are utilized for the Yukawa couplings with matters 
 ${\bf 16}_i~(i=\mbox{generation})$.
A remarkable feature of the model is its high predictivity 
 of the neutrino oscillation parameters 
 as well as reproducing charged fermion masses and mixing angles.

\subsection{Minimal supersymmetric SO(10) model} 
First we give a brief review of the renormalizable minimal SUSY SO(10) model. 
This model was first applied to neutrino oscillation in Ref. 
[\refcite{Babu:1992ia}].
However it did not reproduce the large mixing angles.
It has been pointed out that 
 CP-phases in the Yukawa sector play an important role 
 to reproduce the neutrino oscillation data [\refcite{Matsuda:2000zp}]. 
More detailed analysis incorporating the renormalization group (RG) 
 effects in the context of MSSM [\refcite{Fukuyama:2002ch}]
 has explicitly shown that the model is consistent with the neutrino
 oscillation data at that time and became a realistic model. 
We give a brief review of this renormalizable minimal SUSY SO(10) model.
Yukawa coupling is given by
\begin{eqnarray}
 W_Y = Y_{10}^{ij} {\bf 16}_i H_{10} {\bf 16}_j 
           +Y_{126}^{ij} {\bf 16}_i H_{126} {\bf 16}_j \; , 
\label{Yukawa1}
\end{eqnarray} 
where ${\bf 16}_i$ is the matter multiplet of the $i$-th generation,  
 $H_{10}$ and $H_{126}$ are the Higgs multiplet 
 of {\bf 10} and $\overline{\bf 126} $ representations 
 under SO(10), respectively. 
Note that, by virtue of the gauge symmetry, 
 the Yukawa couplings, $Y_{10}$ and $Y_{126}$, 
 are, in general, complex symmetric $3 \times 3$ matrices. 
After the symmetry breaking pattern of SO(10) to 
${\rm SU}(3)_c \times {\rm SU}(2)_L \times {\rm U}(1)_Y$
via ${\rm SU}(4)_c \times {\rm SU}(2)_L \times {\rm SU}(2)_R$
or ${\rm SU}(5) \times {\rm U}(1)$,
 we find two pair of Higgs doublets 
 in the same representation as the pair in the MSSM. 
One pair comes from $({\bf 1},{\bf 2},{\bf 2}) \subset {\bf 10}$ 
 and the other comes from 
 $({\bf 15}, {\bf 2}, {\bf 2}) \subset \overline{\bf 126}$. 
Using these two pairs of the Higgs doublets, 
 the Yukawa couplings of Eq.~(\ref{Yukawa1}) are rewritten as 
\begin{eqnarray}
W_Y &=& (U^c)_i  \left(
Y_{10}^{ij}  H^u_{10} + Y_{126}^{ij}  H^u_{126} \right) Q
+ (D^c)_i  \left(
Y_{10}^{ij}  H^d_{10} + Y_{126}^{ij}  H^d_{126} \right) Q_j  
\nonumber \\ 
&+& (N^c)_i \left( 
Y_{10}^{ij}  H^u_{10} - 3 Y_{126}^{ij} H^u_{126} \right) L_j 
+ (E^c)_i  \left(
Y_{10}^{ij}  H^d_{10}  - 3 Y_{126}^{ij} H^d_{126} \right) L_j   
\nonumber \\
&+&
 L_i \left( Y_{126}^{ij} \; v_T \right) L_j +
(N^c)_i \left( Y_{126}^{ij} \; v_R \right) (N^c)_j \;  , 
\label{Yukawa2}
\
\end{eqnarray} 
where $U^c$, $D^c$, $N^c$ and 
 $E^c$ are the right-handed ${\rm SU}(2)_L$ 
 singlet quark and lepton superfields, $Q$ and $L$ 
 are the left-handed ${\rm SU}(2)_L$ doublet quark and lepton superfields, 
 $H_{10}^{u,d}$ and $H_{126}^{u,d}$ 
 are up-type and down-type Higgs doublet superfields 
 originated from $H_{10}$ and $H_{126}$, respectively, 
 and the last term is the Majorana mass term 
 of the right-handed neutrinos developed 
 by the vacuum expectation value (VEV)
of the $({\bf 10}, {\bf 1}, {\bf 3})$ Higgs, $v_R$. 
The factor $-3$ in the lepton sector 
 is the Clebsch-Gordan (CG) coefficient. 

In order to preserve the successful gauge coupling unification, 
 suppose that one pair of Higgs doublets 
 given by a linear combination $H_{10}^{u,d}$ and $H_{126}^{u,d}$ 
 is light while the other pair is  heavy ($\geq M_{\rm GUT}$).  
The light Higgs doublets are identified as 
 the MSSM Higgs doublets ($H_u$ and $H_d$) 
 and given by 
\begin{eqnarray} 
 H_u &=& \tilde{\alpha}_u  H_{10}^u  
      + \tilde{\beta}_u  H_{126}^u \;,
 \nonumber \\
 H_d &=& \tilde{\alpha}_d  H_{10}^d  
      + \tilde{\beta}_d  H_{126}^d  \; , 
 \label{mix}
\end{eqnarray} 
where $\tilde{\alpha}_{u,d}$ and $\tilde{\beta}_{u,d}$ 
 denote elements of the unitary matrix  
 which rotate the flavor basis in the original model 
 into the (SUSY) mass eigenstates. 
Omitting the heavy Higgs mass eigenstates, 
 the low energy superpotential is described 
 by only the light Higgs doublets $H_u$ and $H_d$ such that 
\begin{eqnarray}
W_Y &=& 
(U^c) _i \left( \alpha^u  Y_{10}^{ij} + 
\beta^u Y_{126}^{ij} \right)  H_u \, Q_j 
+ (D^c)_i  
\left( \alpha^d  Y_{10}^{ij} + 
\beta^d Y_{126}^{ij}  \right) H_d \,Q_j  \nonumber \\ 
&+& (N^c)_i  
\left( \alpha^u  Y_{10}^{ij} -3 
\beta^u Y_{126}^{ij} \right)  H_u \,L_j 
+ (E^c)_i  
\left( \alpha^d  Y_{10}^{ij} -3 
\beta^d  Y_{126}^{ij}  \right) H_d \,L_j \nonumber \\ 
&+& 
  L_i \left( Y_{126}^{ij} \; v_T \right) L_j + 
 (N^c)_i  
  \left( Y_{126}^{ij} v_R \right)  (N^c)_j \; ,  
\label{Yukawa3}
\end{eqnarray}
where the formulas of the inverse unitary transformation 
 of Eq.~(\ref{mix}), 
 $H_{10}^{u,d} = \alpha^{u,d} H_{u,d} + \cdots $ and 
 $H_{126}^{u,d} = \beta^{u,d} H_{u,d} + \cdots $, 
 have been used. 
Note that the elements of the unitary matrix, 
 $\alpha^{u,d}$ and $\beta^{u,d}$,   
 are in general complex parameters, 
 through which CP-violating phases are introduced 
 into the fermion mass matrices. 

Providing the Higgs VEVs, 
 $H_u = v \sin \beta$ and $H_d = v \cos \beta$ 
 with $v=174 \mbox{GeV}$, 
 the quark and lepton mass matrices can be read off as%
\begin{eqnarray}
  M_u &=& c_{10} M_{10} + c_{126} M_{126}   \nonumber \\
  M_d &=&     M_{10} +     M_{126}   \nonumber \\
  M_D &=& c_{10} M_{10} -3 c_{126} M_{126}   \nonumber \\
  M_e &=&     M_{10} -3     M_{126}   \nonumber \\
  M_T &=& c_T M_{126} \nonumber \\  
  M_R &=& c_R M_{126}  \; , 
 \label{massmatrix}
\end{eqnarray} 
where $M_u$, $M_d$, $M_D$, $M_e$, $M_T$, and $M_R$ 
 denote the up-type quark, down-type quark, 
 Dirac neutrino, charged-lepton, left-handed Majorana, and 
 right-handed Majorana neutrino mass matrices, respectively. 
Note that all the quark and lepton mass matrices 
 are characterized by only two basic mass matrices, $M_{10}$ and $M_{126}$,   
 and four complex coefficients 
 $c_{10}$, $c_{126}$, $c_T$ and $c_R$, 
 which are defined as 
 $M_{10}= Y_{10} \alpha^d v \cos\beta$, 
 $M_{126} = Y_{126} \beta^d v \cos\beta$, 
 $c_{10}= (\alpha^u/\alpha^d) \tan \beta$, 
 $c_{126}= (\beta^u/\beta^d) \tan \beta $, 
 $c_T = v_T/( \beta^d  v  \cos \beta)$) and 
 $c_R = v_R/( \beta^d  v  \cos \beta)$), respectively.  
These are the mass matrix relations required by 
 the minimal SO(10) model. 
In the following in Part I we set $c_T=0$ as the first approximation.
Except for $c_R$, 
  which is used to determine the overall neutrino mass scale, 
 this system has fourteen free parameters in total [\refcite{Matsuda:2000zp}], 
which are fixed from thirteen experimental data of quarks and charged leptons
leaving only one parameter $\sigma$ undetermined. 
(See Ref. [\refcite{Fukuyama:2002ch}] for the definition of $\sigma$.)
Then we can fit all the neutrino oscillation data by fitting $\sigma$ and $c_R$.
The reasonable results found in Ref. [\refcite{Fukuyama:2002ch}] 
are listed in Table 1.
\begin{table}[pt]
\tbl{The input values of $\tan \beta$, $m_s(M_Z)$
and $\delta$ in the CKM matrix and the outputs for the neutrino
oscillation parameters.}
{\begin{tabular}{c|cc|c|ccc|c}
\hline \hline
 $\tan \beta $ & $m_s(M_Z)$ & $\delta$  & $\sigma $
 & $\sin^2 2 \theta_{1 2}$
 & $\sin^2 2 \theta_{2 3}$
 & $\sin^2 2 \theta_{1 3} $
 & $\Delta m_{\odot}^2/\Delta m_{\oplus}^2$ \\ \hline
40 & 0.0718 & $ 93.6^\circ $ & 3.190& 
0.738 & 0.900 & 0.163 & 0.205 \\
45 & 0.0729 & $ 86.4^\circ $ & 3.198& 
0.723 & 0.895 & 0.164 & 0.188 \\
50 & 0.0747 & $ 77.4^\circ $ & 3.200& 
0.683 & 0.901 & 0.164 & 0.200 \\
55 & 0.0800 & $ 57.6^\circ $ & 3.201& 
0.638 & 0.878 & 0.152 & 0.198 \\
\hline \hline
\end{tabular}}
\end{table}

\begin{figure}[th]
\centerline{\psfig{file=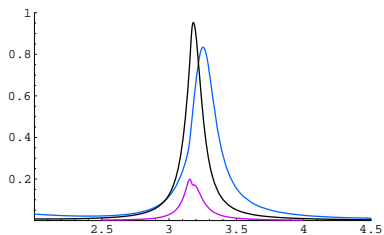,width=5cm}}
\vspace*{8pt}
\caption{Three mixing angles in the MNS matrix as functions
of $\sigma \mbox{[rad]}$. 
The graphs with the highest, middle and lowest peaks
correspond to $\sin^2 2 \theta_{2 3}$, 
$\sin^2 2 \theta_{1 2}$ and $\sin^2 2 \theta_{1 3}$, respectively. 
The plots of $\sin^2 2 \theta_{23}$ and $\sin^2 2 \theta_{13}$ have 
the sharp peaks at $\sigma \sim 3.2 [\mbox{rad}]$, 
while $\sin^2 2 \theta_{12}$ has the sharp peak at 
$\sigma \sim 3.3 [\mbox{rad}]$ cited from Ref. [4].}
\end{figure}

As mentioned above, 
 our resultant neutrino oscillation parameters 
 are sensitive to all the input parameters. 
In other words, if we use the neutrino oscillation data 
 as the input parameters, 
 the other input, for example, the CP-phase in the CKM matrix 
 can be regarded as the prediction of our model. 
It is a very interesting observation 
 that the CP-phases listed above are 
 in the region consistent with experiments.
The CP-violation in the lepton sector
 is characterized by the Jarlskog parameter
 defined as 
\begin{eqnarray}
 J_{CP} = \mbox{Im}\left[ 
     U_{e2} U_{\mu 2}^*  U_{e 3}^*  U_{\mu 3} \right] \; , 
\end{eqnarray}  
where $U_{f i}$ is the MNS matrix element. 

It is well known that the SO(10) GUT model possesses 
 a simple mechanism of baryogenesis 
 through the out-of-equilibrium decay 
 of the right-handed neutrinos, 
 namely, the leptogenesis [\refcite{Fukugita-Yanagida}]. 
The lepton asymmetry in the universe is generated by CP-violating 
out-of-equilibrium decay of the heavy neutrinos,  
$N \rightarrow \ell_L H_u^*$ and $N \rightarrow \overline{\ell_L} H_u$. 
The leading contribution is given by the interference between 
the tree level and one-loop level decay amplitudes, 
and the CP-violating parameter is found to be
\begin{eqnarray}
\epsilon = 
\frac{1}{8 \pi (Y_\nu Y_\nu^\dag)_{11}}
\sum_{j=2,3}\mbox{Im} \left[ (Y_\nu Y_\nu^\dag)_{1j}^2 \right]
\left\{ f(M_{Rj}^2/M_{R1}^2)
+ 2 g(M_{Rj}^2/M_{R1}^2) \right\} \; .
\label{epsilon}
\end{eqnarray}
Here $f(x)$ and $g(x)$ correspond to 
the vertex and the wave function corrections, 
\begin{eqnarray}
f(x)&\equiv& \sqrt{x} \left[
1-(1+x)\mbox{ln} \left(\frac{1+x}{x} \right) \right] \;,
\nonumber\\
g(x)&\equiv& \frac{\sqrt{x}}{2(1-x)}   \; ,  
\end{eqnarray}
respectively, and both are reduced to 
$\sim -\frac{1}{2 \sqrt{x}}$ for $ x \gg 1$. 
So in this approximation, $\epsilon$ becomes 
\begin{equation}
\epsilon = - 
\frac{3}{16 \pi (Y_\nu Y_\nu^\dag)_{11}}
\sum_{j=2,3} \mbox{Im} \left[(Y_\nu Y_\nu^\dag)_{1j}^2 \right]
\frac{M_{R1}}{M_{Rj}}\;.
\end{equation}
These quantities are evaluated by using the results 
 presented in Table 1, 
 and the results are listed in Table 2.  
\begin{table}[pt]
\tbl{The input values of $\tan \beta$ and the outputs 
for the CP-violating observables}
{\begin{tabular}{c|ccc}
\hline \hline
 $\tan \beta $ & 
 $ \langle m_\nu \rangle_{ee}~ (\mbox{eV})$ & 
 $J_{CP}$ &   $\epsilon$  \\   \hline
40 & 0.00122  & $~~0.00110$ & $ 7.39 \times 10^{-5} $ \\
45 & 0.00118  & $-0.00429$  & $ 6.80 \times 10^{-5} $ \\
50 & 0.00119  & $-0.00631$  & $ 6.50 \times 10^{-5} $ \\
55 & 0.00117  & $-0.00612$  & $ 11.2 \times 10^{-5} $ \\ 
\hline \hline
\end{tabular}}
\end{table}

Now we turn to the discussion about the rate of 
the lepton flavor violating (LFV) processes
and the muon $g-2$.
The evidence of the neutrino flavor mixing implies that 
 the lepton flavor of each generation is not individually conserved. 
Therefore the LFV processes 
 in the charged-lepton sector such as 
 $\mu \rightarrow e \gamma$, $\tau \rightarrow \mu \gamma$ 
 are allowed. 
In simply extended models 
 so as to incorporate massive neutrinos into the standard model, 
 the rate of the LFV processes is accompanied 
 by a highly suppression factor, 
 the ratio of neutrino mass to the weak boson mass, 
 because of the GIM mechanism,  
 and is far out of the reach of the experimental detection. 
However, in supersymmetric models, the situation is quite different. 
In this case, soft SUSY breaking parameters can be new LFV sources, 
 and the rate of the LFV processes 
 are suppressed by only the scale 
 of the soft SUSY breaking parameters 
 which is assumed to be the electroweak scale. 
Thus the huge enhancement occurs compared to the previous case. 
In fact, the LFV processes can be one of the most important processes 
 as the low-energy SUSY search. 
we evaluate the rate of the LFV processes 
 in the minimal SUSY SO(10) model [\refcite{fukuyama2}], 
 where the neutrino Dirac Yukawa couplings 
 are the primary LFV sources. 
Although in Ref. [\refcite{Fukuyama:2002ch}] 
 various cases with given $\tan \beta = 40-55$ have been analyzed, 
 we consider only the case $\tan \beta =45$ in the following. 
Our final result in the next section is almost insensitive 
 to $\tan \beta$ values in the above range. 
The predictions of the minimal SUSY SO(10) model 
 necessary for the LFV processes are as follows [\refcite{Fukuyama:2002ch}]: 
 with $\sigma=3.198$ fixed, 
 the right-handed Majorana neutrino mass eigenvalues 
 are found to be (in GeV) 
 $M_{R_1}=1.64 \times 10^{11}$,  
 $M_{R_2}=2.50 \times 10^{12}$ and 
 $M_{R_3}=8.22 \times 10^{12}$, 
 where $c_R$ is fixed so that 
 $\Delta m_\oplus^2 = 2 \times 10^{-3} \mbox{eV}^2$. 
In the basis where both of the charged-lepton 
 and right-handed Majorana neutrino mass matrices 
 are diagonal with real and positive eigenvalues, 
 the neutrino Dirac Yukawa coupling matrix at the GUT scale 
 is found to be 
\begin{eqnarray}
 Y_{\nu} = 
\left( 
 \begin{array}{ccc}
-0.000135 - 0.00273 i & 0.00113  + 0.0136 i  & 0.0339   + 0.0580 i  \\ 
 0.00759  + 0.0119 i  & -0.0270   - 0.00419  i  & -0.272    - 0.175   i  \\ 
-0.0280   + 0.00397 i & 0.0635   - 0.0119 i  &  0.491  - 0.526 i 
 \end{array}   \right) \; .  
\label{Ynu}
\end{eqnarray}     
LFV effect most directly emerges 
 in the left-handed slepton mass matrix 
 through the RGEs such as [\refcite{Hisano}]:
\begin{eqnarray}
\mu \frac{d}{d \mu} 
  \left( m^2_{\tilde{\ell}} \right)_{ij}
&=&  \mu \frac{d}{d \mu} 
  \left( m^2_{\tilde{\ell}} \right)_{ij} \Big|_{\mbox{MSSM}} 
 \nonumber \\
&+& \frac{1}{16 \pi^2} 
\left( m^2_{\tilde{\ell}} Y_{\nu}^{\dagger} Y_{\nu}
 + Y_{\nu}^{\dagger} Y_{\nu} m^2_{\tilde{\ell}} 
 + 2  Y_{\nu}^{\dagger} m^2_{\tilde{\nu}} Y_{\nu}
 + 2 m_{H_u}^2 Y_{\nu}^{\dagger} Y_{\nu} 
 + 2  A_{\nu}^{\dagger} A_{\nu} \right)_{ij}  \; ,
 \label{RGE} 
\nonumber\\
\end{eqnarray}
where the first term in the right hand side denotes 
 the normal MSSM term with no LFV. 
We have found $Y_\nu$ explicitly and we can calculate LFV and
 related phenomena unambiguously [\refcite{fukuyama2}]
In the leading-logarithmic approximation, 
 the off-diagonal components ($i \neq j$)
 of the left-handed slepton mass matrix are estimated as 
\begin{eqnarray}
 \left(\Delta  m^2_{\tilde{\ell}} \right)_{ij}
 \sim - \frac{3 m_0^2 + A_0^2}{8 \pi^2} 
 \left( Y_{\nu}^{\dagger} L Y_{\nu} \right)_{ij} \; ,  
 \label{leading}
\end{eqnarray}
where the distinct thresholds of the right-handed 
 Majorana neutrinos are taken into account 
 by the matrix $ L = \log [M_{\rm GUT}/M_{R_i}] \delta_{ij}$. 

The effective Lagrangian 
 relevant for the LFV processes ($\ell_i \rightarrow \ell_j \gamma$) 
 and the muon $g-2$ is described as 
\begin{eqnarray}
 {\cal L}_{\mbox{eff}}= 
 -  \frac{e}{2} m_{\ell_i} \overline{\ell}_j \sigma_{\mu \nu} F^{\mu \nu} 
 \left(A_L^{j i} P_L + A_R^{j i} P_R \right) \ell_i  \; , 
\end{eqnarray}
where $P_{R, L} = (1 \pm \gamma_5)/2 $ is  
 the chirality projection operator, 
 and  $A_{L,R}$ are the photon-penguin couplings of 1-loop diagrams 
 in which chargino-sneutrino and neutralino-charged slepton 
 are running. 
The explicit formulas of $A_{L,R}$ etc. used in our analysis 
 are summarized in Ref. [\refcite{Hisano-etal}, \refcite{Okada-etal}].
The rate of the LFV decay of charged-leptons is given by 
\begin{eqnarray}
\Gamma (\ell_i \rightarrow \ell_j \gamma) 
= \frac{e^2}{16 \pi} m_{\ell_i}^5 
 \left( |A_L^{j i}|^2  +  |A_R^{j i}|^2  \right) \; , 
\end{eqnarray}
while the real diagonal components of $A_{L,R}$ 
 contribute to the anomalous magnetic moments of 
 the charged-leptons such as 
\begin{eqnarray}
 \delta a_{\ell_i}^{\rm SUSY} = \frac{g_{\ell_i}-2}{2} 
  = -  m_{\ell_i}^2 
  \mbox{Re} \left[ A_L^{i i}  +  A_R^{i i}  \right]  \; . 
\end{eqnarray}
In order to clarify the parameter dependence 
 of the decay amplitude, 
 we give here an approximate formula of the LFV decay rate 
 [\refcite{Hisano-etal}], 
\begin{eqnarray}
\Gamma (\ell_i \rightarrow \ell_j \gamma) 
 \sim  \frac{e^2}{16 \pi} m_{\ell_i}^5 
 \times  \frac{\alpha_2}{16 \pi^2} 
 \frac{\left| (\Delta  m^2_{\tilde{\ell}} )_{ij}\right|^2}{M_S^8} 
 \tan^2 \beta \; , 
 \label{LFVrough}
\end{eqnarray}
where $M_S$ is the average slepton mass at the electroweak scale, 
 and $ \left(\Delta  m^2_{\tilde{\ell}} \right)_{ij}$ 
 is the slepton mass estimated in Eq.~(\ref{leading}). 
We can see that the neutrino Dirac Yukawa coupling matrix 
 plays the crucial role in calculations of the LFV processes. 
We use the neutrino Dirac Yukawa coupling matrix of Eq.~(\ref{Ynu})
 in our numerical calculations. 

\begin{figure}[th]
\begin{center}
\subfigure[]{\includegraphics[width=5.5cm]{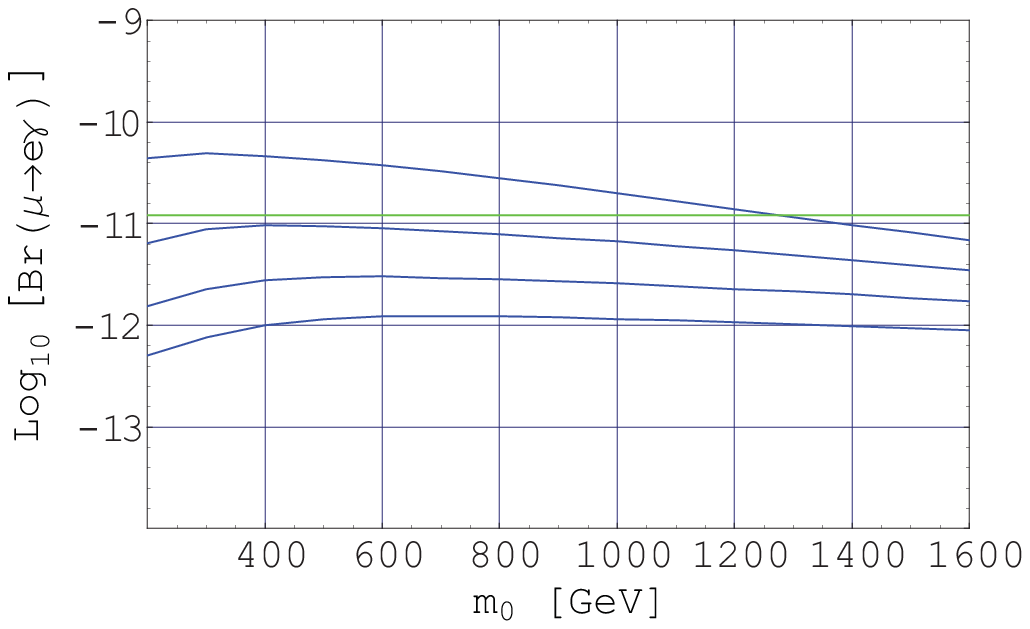}\label{Fig1a}}
\subfigure[]{\includegraphics[width=5.5cm]{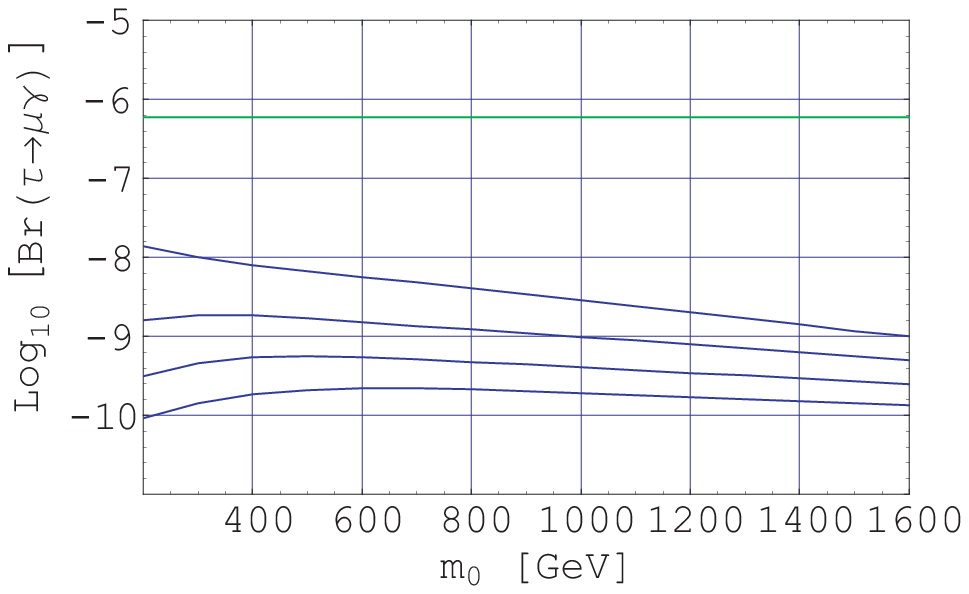}\label{Fig1b}}
\begin{center}
\subfigure[]{\includegraphics[width=5.5cm]{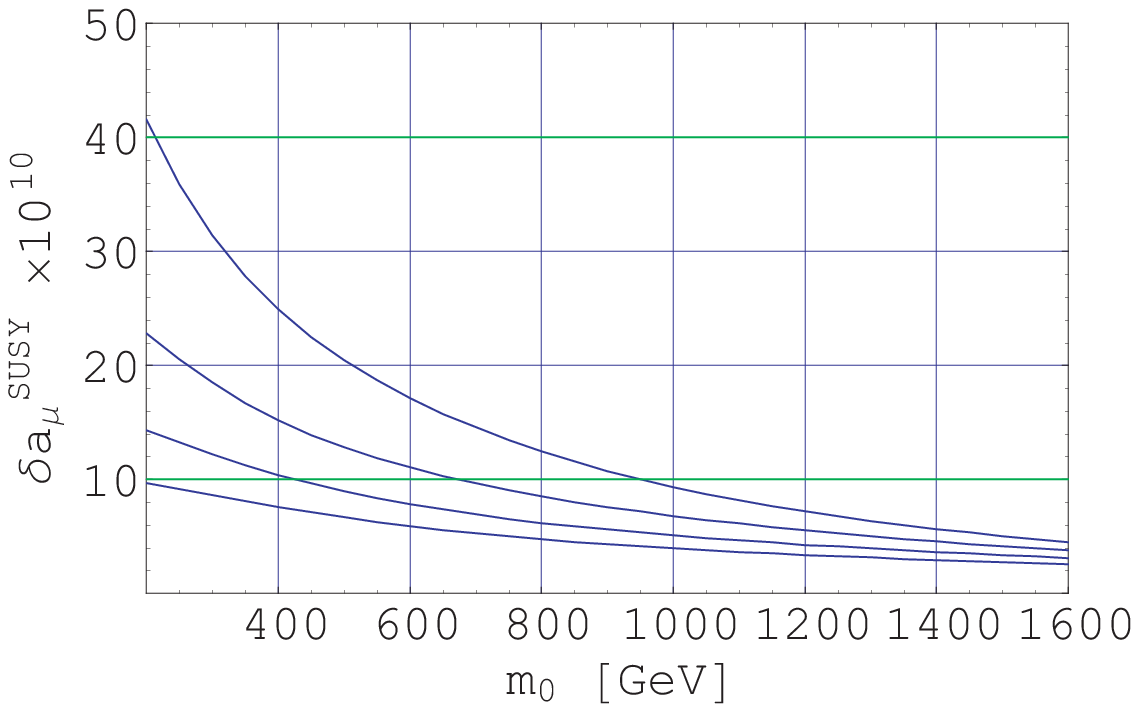}\label{Fig1c}}
\end{center}
\vspace*{8pt}
\caption{
The branching ratios,
(a)
$\mbox{Log}_{10} \left[\mbox{Br}(\mu \rightarrow e \gamma) \right]$, 
(b)
$\mbox{Log}_{10} \left[\mbox{Br}(\tau \rightarrow \mu \gamma)\right]$.
and (c)
the SUSY contribution to the muon $g-2$ in units of $10^{-10}$,
$\delta a_{\ell_i}^{\rm SUSY} = \frac{g_{\ell_i}-2}{2}$.
All these figures are plotted as a function of $m_0$ (GeV) 
 for $M_{1/2} =400, 600, 800, 1000$ GeV (from top to bottom) 
 with $A_0=0$ and $\mu >0$ cited from Ref. [6].}
\end{center}
\end{figure}

\begin{figure}[th]
\begin{center}
\subfigure[]{\includegraphics[width=5.5cm]{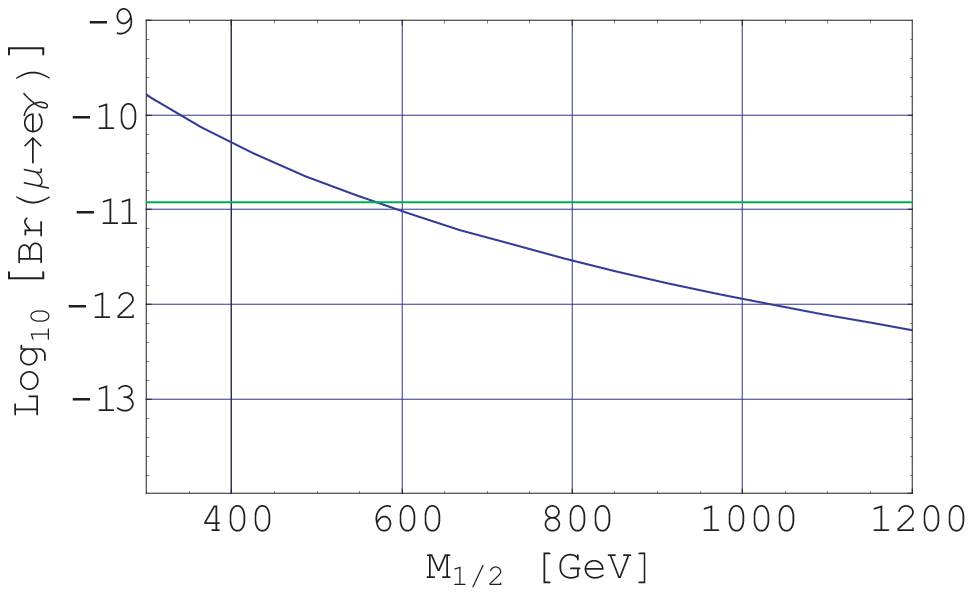}\label{Fig2a}}
\subfigure[]{\includegraphics[width=5.5cm]{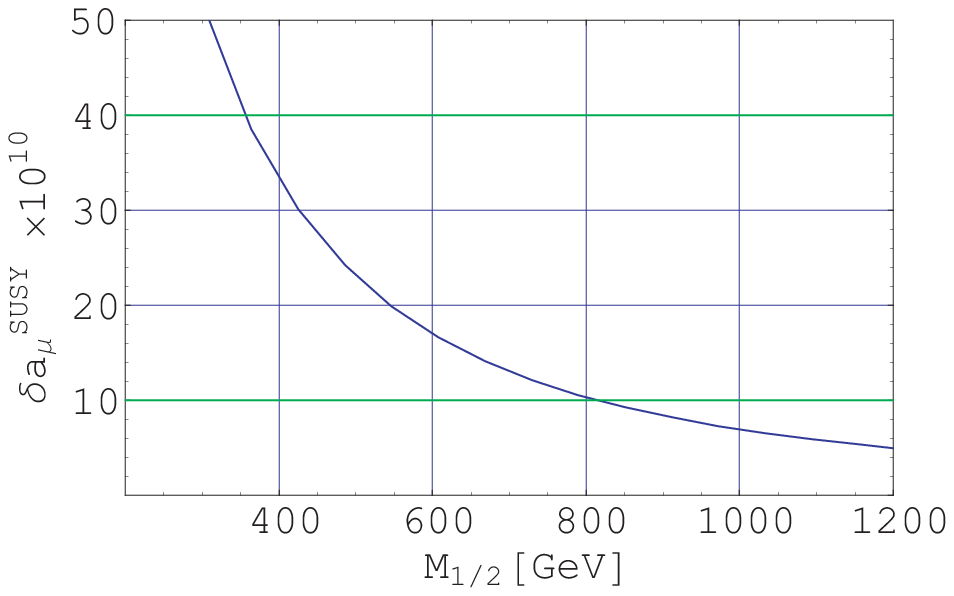}\label{Fig2b}}
\begin{center}
\subfigure[]{\includegraphics[width=5.5cm]{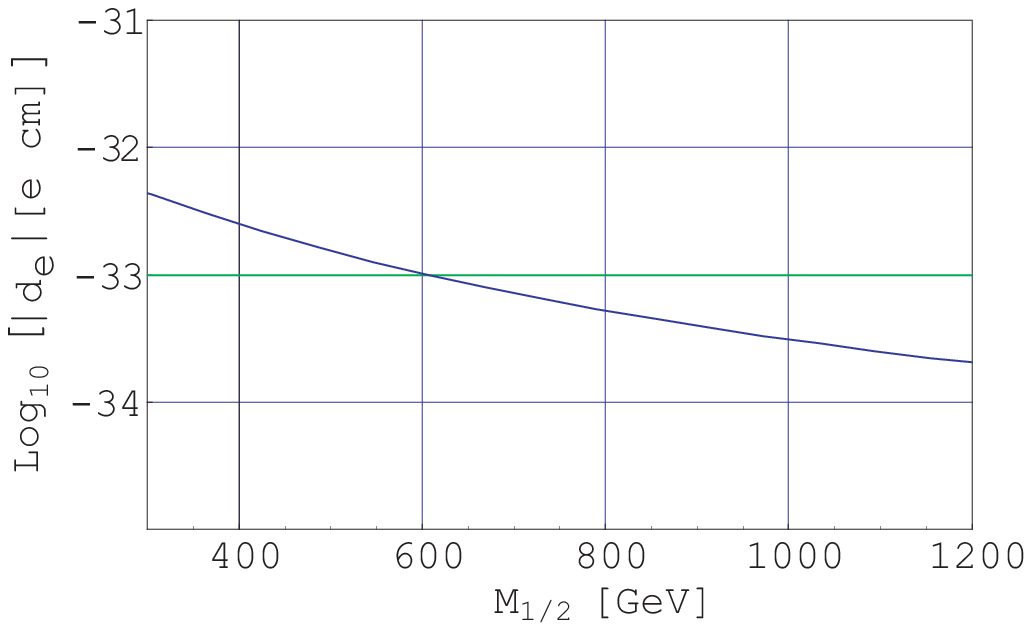}\label{Fig2c}}
\end{center}
\vspace*{8pt}
\caption{
The branching ratio,
(a)
$\mbox{Log}_{10} \left[{\rm Br} (\mu \rightarrow e \gamma) \right]$,
(b) the SUSY contribution to the muon $g-2$ in units of $10^{-10}$, 
$\delta a_{\ell_i}^{\rm SUSY} = \frac{g_{\ell_i}-2}{2}$,
and (c) the electron EDM, 
$\mbox{Log}_{10} \left[ | d_e | [ \mbox{e cm}] \right]$. 
All these figures are plotted as a function of $M_{1/2}$ (GeV) 
 along the cosmological constraint of Eq.~(\ref{relation})
 cited from Ref. [6].}
\end{center}
\end{figure}

The recent Wilkinson Microwave Anisotropy Probe (WMAP) satellite data 
 [\refcite{WMAP}] 
 provide estimations of various cosmological parameters 
 with greater accuracy. 
The current density of the universe is composed of 
 about 73\% of dark energy and 27\% of matter. 
Most of the matter density is in the form of 
 the CDM, and its density is estimated to be (in 2$\sigma$ range) 
\begin{eqnarray}
\Omega_{\rm CDM} h^2 = 0.1126^{+0.0161}_{-0.0181} \; . 
 \label{WMAP} 
\end{eqnarray}
The parameter space of the CMSSM 
 which allows the neutralino relic density 
 suitable for the cold dark matter 
 has been recently re-examined 
 in the light of the WMAP data [\refcite{CDM}]. 
It has been shown that the resultant parameter space 
 is dramatically reduced into the narrow stripe 
 due to the great accuracy of the WMAP data. 
It is interesting to combine this result 
 with our analysis of the LFV processes and the muon $g-2$. 
In the case relevant for our analysis, 
 $\tan \beta=45$, $\mu>0$ and $A_0=0$, 
 we can read off the approximate relation 
 between $m_0$ and $M_{1/2}$ 
 such as (see Figure 1 in the second paper of Ref. [\refcite{CDM}].) 
\begin{eqnarray}
m_0(\mbox{GeV}) = \frac{9}{28} M_{1/2}(\mbox{GeV}) + 150(\mbox{GeV}) \;,  
 \label{relation} 
\end{eqnarray} 
along which the neutralino CDM is realized. 
$M_{1/2}$ parameter space is constrained within the range 
 $300 \mbox{GeV} \leq M_{1/2} \leq 1000 \mbox{GeV}$ 
 due to the experimental bound on the SUSY contribution 
 to the $ b \rightarrow s \gamma$ branching ratio 
 and the unwanted stau LSP parameter region. 
We show $\mbox{Br}(\mu \rightarrow e \gamma)$ and 
 the muon $g-2$ as functions of $M_{1/2}$ 
 in Fig.~\ref{Fig2a} and \ref{Fig2b}, respectively, 
 along the neutralino CDM condition of Eq.~(\ref{relation}). 
We find the parameter region, 
 $560 \mbox{GeV} \leq M_{1/2} \leq 800 \mbox{GeV}$, 
 being consistent with all the experimental data. 

There are a variety of other applications in this model:
The semileptonic flavor violation processes were considered
in Ref. [\refcite{fukuyama3}], for instance,
$\tau^- \to \ell^- \,M\; (M=\pi^0,~\eta,~\eta^\prime,\cdots)$,
$\tau^- \to \ell^- \,V\; (V=\rho^0,~\phi,~\omega,\cdots)$.
The (transition) magnetic moments of the Majorana neutrino 
in the MSSM were first considered in Ref. [\refcite{NMM}],
and found to be an order of magnitude larger than those 
calculated in the standard model extended to incorporate 
the see-saw mechanism. 
However, they are still too small to make the spin flavor precession
in the solar and in the supernovae observable.

When the KamLAND data [\refcite{Eguchi:2002dm}] was released, 
 the results in Ref. [\refcite{Fukuyama:2002ch}] were found to be deviated 
 by 3$\sigma$ from the observations. 
Afterward this minimal SO(10) was modified by many authors, 
 using the so-called type-II see-saw mechanism [\refcite{Bajc:2002iw}] 
($c_T \neq 0$)
 and/or considering a ${\bf 120}$ Higgs coupling to the matter
 in addition to the ${\bf \overline{126}}$ Higgs [\refcite{Goh:2003hf}]. 
Based on an elaborate input data scan [\refcite{Babu:2005ia}, 
\refcite{Bertolini:2006pe}],
 it has been shown that the minimal SO(10) is essentially consistent 
 with low energy data of fermion masses and mixing angles.

On the other hand, it has been long expected to construct 
 a concrete Higgs sector of the minimal SO(10) model. 
The simplest Higgs superpotential at the renormalizable level
is given by [\refcite{clark}, \refcite{lee}, \refcite{aulakh}]
\be
W=m_1 \Phi^2 + m_2 \Delta \overline{\Delta} 
+m_3 H^2
+\lambda_1 \Phi^3 + \lambda_2 \Phi \Delta \overline{\Delta}
+\lambda_3 \Phi \Delta H + \lambda_4 \Phi \overline{\Delta} H \;,
\label{lee}
\ee
where $\Phi ={\bf 210}$, $\Delta ={\bf 126}$, 
$\overline{\Delta} ={\bf \overline{126}}$ and $H={\bf 10}$.
The interactions of ${\bf 210}$, ${\bf \overline{126}}$, 
${\bf 126}$ and ${\bf 10}$ lead to some complexities 
in decomposing the GUT representations to the MSSM 
and in getting the low energy mass spectra.  
Particularly, the CG coefficients 
corresponding to the decompositions of 
${\rm SO}(10) \to 
{\rm SU}(3)_C \times {\rm SU}(2)_L \times {\rm U}(1)_Y$ 
have to be found.
This problem was first attacked by X.~G.~He and S.~Meljanac
[\refcite{he}] and further by D.~G.~Lee [\refcite{lee}] 
and by J.~Sato [\refcite{Joe}]. 
But they did not present the explicit form of mass matrices for 
a variety of Higgs fields and did not perform a formulation 
of the proton decay analysis.  
We completed that program in [\refcite{Fukuyama:2004xs}, 
\refcite{Bajc:2004xe}, \refcite{Aulakh:2004hm}].
This construction gives some constraints among the VEVs 
of several Higgs multiplets, 
 which give rise to a trouble in the gauge coupling unification 
 [\refcite{Bertolini:2006pe}]. 
The importance of the threshold corrections 
was also discussed in Ref. [\refcite{Parida}].
The trouble comes from the fact that the observed neutrino oscillation
 data suggests the right-handed neutrino mass around $10^{13-14}$ GeV,
 which is far below the GUT scale. 
This intermediate scale is provided by Higgs field VEV, 
 and several Higgs multiplets are expected to have their masses 
 around the intermediate scale and contribute to 
 the running of the gauge couplings. 
Therefore, the gauge coupling unification 
 at the GUT scale may be spoiled. 
This fact has been explicitly shown in Ref. [\refcite{Bertolini:2006pe}], 
 where the gauge couplings are not unified any more 
 and even the ${\rm SU}(2)$ gauge coupling blows up below the GUT scale. 
In order to avoid this trouble and keep the successful gauge coupling 
 unification as usual, it is desirable that all Higgs multiplets 
 have masses around the GUT scale, but some Higgs fields 
 develop VEVs at the intermediate scale. 
More Higgs multiplets and some parameter tuning in the Higgs sector 
 are necessary to realize such a situation. 

In addition to the issue of the gauge coupling unification, 
 the minimal SO(10) model potentially suffers from the problem 
 that the gauge coupling blows up around the GUT scale. 
This is because the model includes many Higgs multiplets of 
 higher dimensional representations. 
In field theoretical point of view, this fact implies 
 that the GUT scale is a cutoff scale of the model, 
 and more fundamental description of the minimal SO(10) model 
 would exist above the GUT scale.

\section{Part II}

In this Part we propose a solution to the problem of the minimal SO(10)
discussed in Part I. \footnote{This part is based on the work of Ref.
[\refcite{fko}]:
``Solving problems of 4D minimal SO(10) model in a warped extra dimension'',
T. Fukuyama, T. Kikuchi and N. Okada,
e-Print Archive: hep-ph/0702048}

\subsection{Minimal SO(10) model in a warped extra dimension} 
 We consider the minimal SUSY SO(10) model 
in the following 5D warped geometry [\refcite{RS}],
\begin{eqnarray}
 d s^2 = e^{-2 k r_c |y|} \eta_{\mu \nu} d x^{\mu} d x^{\nu} 
 - r_c^2 d y^2 \; , 
\end{eqnarray}
 for $-\pi\leq y\leq\pi$, where $k$ is the AdS curvature, and 
 $r_c$ and $y$ are the radius and the angle of $S^1$, respectively. 
The most important feature of the warped extra dimension model 
 is that the mass scale of the IR brane is warped down to 
 a low scale by the warp factor [\refcite{RS}], $ \omega \equiv e^{-k r_c \pi}$, 
 in four dimensional effective theory. 
For simplicity, we take the cutoff of the original five dimensional theory 
 and the AdS curvature as $M_5 \simeq k \simeq M_P$, 
 the four dimensional Planck mass,
 and so we obtain the effective cutoff scale 
 as $\Lambda_{\rm IR}= \omega M_P$ in effective four dimensional theory. 
Now let us take the warp factor so as for the GUT scale 
 to be the effective cutoff scale 
 $ M_{\rm GUT}= \Lambda_{\rm IR}=\omega M_P$ [\refcite{Nomura:2006pn}]. 
As a result, we can realize, as four dimensional effective theory, 
 the minimal SUSY SO(10) model 
 with the effective cutoff at the GUT scale. 

Before going to a concrete setup of the minimal SO(10) model 
 in the warped extra dimension, 
 let us see Lagrangian for the hypermultiplet in the bulk, 
\begin{eqnarray}
{\cal L} &=& \int dy \left\{ 
\int d^4 \theta \; r_c \; e^{- 2 k r_c |y|} 
 \left( 
 H^{\dagger} e^{- V} H + H^{c} e^{ V}H^{c \dagger} 
 \right) \right. \nonumber \\
&+& 
\left. 
\int d^2 \theta e^{-3 k r_c |y|}
 H^{c} \left[ 
  \partial_{y} - \left( \frac{3}{2}-c \right) k r_c \epsilon(y) 
 - \frac{\chi}{\sqrt{2}}  \right]  
 H  +h.c. \right \} \; , 
\label{bulkL}
\end{eqnarray}
where $c$ is a dimensionless parameter, 
$\epsilon(y)=y/|y|$ is the step function, 
 $H, ~H^c$ is the hypermultiplet charged under some gauge group, 
 and $V, ~\chi$
 are the vector multiplet and the adjoint chiral multiplets, 
 which form an $N=2$ SUSY gauge multiplet. 
 $Z_2$ parity for $H$ and $V$ is assigned as even, 
 while odd for $H^c$ and $\chi$. 

When the gauge symmetry is broken down, 
 it is generally possible that the adjoint chiral multiplet 
 develops its VEV [\refcite{Kitano:2003cn}]. 
Since its $Z_2$ parity is odd, 
 the VEV has to take the form, 
\bea
\left<\Sigma \right> = 2 \alpha k r_c  \epsilon(y) \; , 
\eea
where the VEV has been parameterized by a parameter $\alpha$. 
In this case, the zero mode wave function of $H$ 
 satisfies the following equation of motion:
\bea
\left[\partial_y -
 \left(\frac{3}{2}-c + \alpha \right) k r_c \epsilon(y) \right]H =0 
\eea
which yields 
\bea
H = \frac{1}{\sqrt{N}} 
 e^{ (3/2-c + \alpha) kr_c |y|} \; h(x^\mu) \; , 
\eea
where $h(x^\mu)$ is the chiral multiplet in four dimensions. 
Here, $N$ is a normalization constant 
 by which the kinetic term is canonically normalized, 
\be
\frac{1}{N} 
 =  \frac{(1-2 c+2 \alpha) k }
{e^{(1-2 c+2 \alpha) k r_c \pi}-1} \; . 
\ee
Lagrangian for a chiral multiplets on the IR brane is given by 
\bea 
 {\cal L}_{\rm IR}=  
 \int d^4 \theta \;  \omega^\dagger \omega \;  \Phi^\dagger \Phi 
 +\left[   \int d^2 \theta \;  \omega^3 \; W(\Phi) + h.c.   
 \right] \; ,
\eea 
where we have omitted the gauge interaction part 
 for simplicity. 
If it is allowed by the gauge invariance, 
 we can write the interaction term 
 between fields in the bulk and on the IR brane, 
\bea 
{\cal L}_{int}= \int d^2 \theta \omega^3 
 \frac{Y}{\sqrt{M_5}} \Phi^2 H(y=\pi) +h.c. \; ,  
\label{IR-Yukawa} 
\eea  
where $Y$ is a Yukawa coupling constant, 
 and $M_5$ is the five dimensional Planck mass 
 (we take $M_5\sim M_P$ as mentioned above, for simplicity). 
Rescaling the brane field $\Phi \rightarrow  \Phi/\omega$ 
 to get the canonically normalized kinetic term 
 and substituting the zero-mode wave function of the bulk fields, 
 we obtain Yukawa coupling constant 
 in effective four dimensional theory as 
\bea 
  Y_{4D} \sim Y 
\eea 
 if $e^{ (1/2 - c +  \alpha ) k r_c \pi}  \gg 1$,  
 while 
\bea 
  Y_{4D} \sim Y 
  \times e^{ (1/2 -  c +  \alpha ) k r_c \pi}  \ll Y \; , 
\label{suppression}
\eea  
 for $e^{ (1/2 - c + \alpha ) k r_c \pi }  \ll 1$. 
In the latter case, we obtain a suppression factor 
 since $H$ is localized around the UV brane. 

Now we give a simple setup of the minimal SO(10) model 
 in the warped extra dimension. 
We put all ${\bf 16}$ matter multiplets on the IR ($y=\pi$) brane, 
 while the Higgs multiplets ${\bf 10}$ and $\overline{\bf 126}$ 
 are assumed to live in the bulk. 
In Eq.~(\ref{IR-Yukawa}), replacing the brane field into the matter 
 multiplets and the bulk field into the Higgs multiplets, 
 we obtain Yukawa couplings in the minimal SO(10) model. 
The Lagrangian for the bulk Higgs multiplets are given 
 in the same form as Eq.~(\ref{bulkL}), 
 where $\chi$ is the SO(10) adjoint chiral multiplet, ${\bf 45}$. 
As discussed above, since the SO(10) gauge group is broken 
 down to the SM one, 
 some components in $\chi$ which is singlet under the SM gauge group 
 can in general develop VEVs. 
Here we consider a possibility that 
 the ${\rm U}(1)_X$ component 
 in the adjoint $\chi ={\bf 45}$ under the decomposition 
 SO(10) $\supset {\rm SU}(5) \times {\rm U}(1)_X$ has 
 a non-zero VEV,
\bea
 {\bf 45} = {\bf 1}_0
 \oplus {\bf 10}_{+4} \oplus \overline{\bf 10}_{-4}
 \oplus {\bf 24}_0 \; .   \nonumber 
\eea
The $\overline{\bf 126}$ Higgs multiplet 
 are decomposed under ${\rm SU}(5) \times {\rm U}(1)_X$ as
\bea
\overline{\bf 126} &=& 
 {\bf 1}_{+10} 
 \oplus {\bf 5}_{+2} \oplus \overline{\bf 10}_{+6}
 \oplus {\bf 15}_{-6} 
 \oplus \overline{\bf 45}_{-2} \oplus {\bf 50}_{+2} \; . 
 \nonumber  
\eea
In this decomposition, 
 the coupling between a bulk Higgs multiplet and 
 the ${\rm U}(1)_X$ component in $\chi$ is proportional 
 to  ${\rm U}(1)_X$ charge, 
\be
{\cal L}_{int} \supset \frac{1}{2} \int d^2 \theta \omega^3
Q_X \langle \Sigma_X \rangle H^c H + h.c. \;,
\ee
 and thus each component effectively obtains 
 the different bulk mass term,  
\bea 
  \left( \frac{3}{2} - c \right) k r_c  
   + \frac{1}{2}Q_X \langle \Sigma_X \rangle,   
\eea
 where $Q_X$ is the ${\rm U}(1)_X$ charge of corresponding Higgs multiplet, 
 and $\Sigma_X$ is the scalar component of 
 the ${\rm U}(1)_X$ gauge multiplet (${\bf 1}_0$). 
Now we obtain different configurations of the wave functions 
 for these Higgs multiplets. 
Since the ${\bf 1}_{+10}$ Higgs has a large ${\rm U}(1)_X$ charge 
 relative to other Higgs multiplets, 
 we can choose parameters $c$ and $\langle \Sigma_X \rangle$ 
 so that Higgs doublets are mostly localized around the IR brane 
 while the ${\bf 1}_{+10}$ Higgs is localized around the UV brane. 
Therefore, 
 we obtain a suppression factor 
 as in Eq.~(\ref{suppression}) 
 for the effective Yukawa coupling between 
 the Higgs and right-handed neutrinos. 
In effective four dimensional description, 
 the GUT mass matrix relation is partly broken down, 
 and the last term in Eq.~(\ref{Yukawa3}) is replaced into 
\bea 
 Y_{126}^{ij} v_R \rightarrow Y_{126}^{ij} (\epsilon v_R) \; ,  
\eea
where $\epsilon$ denotes the suppression factor. 
By choosing an appropriate parameters 
 so as to give $\epsilon=10^{-2}-10^{-3}$, 
 we can take $v_R \sim M_{\rm GUT}$ 
 and keep the successful gauge coupling unification in the MSSM. 

In our setup, all the matters reside on the brane 
 while the Higgs multiplets reside in the bulk. 
This setup shares the same advantage as 
 the so-called orbifold GUT [\refcite{Kawamura:2000ev},
\refcite{Altarelli:2001qj}, \refcite{Hall:2001pg}]. 
We can assign even $Z_2$ parity 
 for MSSM doublet Higgs superfields 
 while odd for triplet Higgs superfields, 
 as a result, the proton decay process through 
 dimension five operators are forbidden. 
This is especially important for the minimal supersymmetric SO(10) model 
since it gives rather large $\tan \beta$, as was shown in 
Tables 1 and 2, and since the proton decay ratio is proportional
to $\tan^4 \beta$ [\refcite{Goh:2003nv}, \refcite{Fukuyama:2004pb},
\refcite{Dutta:2004zh}].

\subsection{Conclusion}
The minimal renormalizable supersymmetric SO(10) model 
 is a simple framework to reproduce current data 
 for fermion masses and flavor mixings with some predictions. 
Above that it gives full informations of all mass matrices including
those of Dirac neutrino, left-handed and heavy right-handed neutrino
with full phases, unambiguously. This enables us to predict the wide ranges of physics, for instance, neutrinoless double beta decay, LFV, lepton anomalous moments, leptogenesis etc., which gave all consistent with the present data.
However, this model suffers from some problems related to 
 the running of the gauge couplings. 
To fit the neutrino oscillation data, 
 the mass scale of right-handed neutrinos lies 
 at the intermediate scale. 
This implies the presence of some Higgs multiplets 
 lighter than the GUT scale. 
As a result, the gauge coupling unification in the MSSM 
 may be spoiled. 
In addition, since Higgs multiplets of large representations 
 are introduced in the model, 
 the gauge couplings blow up around the GUT scale. 
Thus, the minimal SO(10) model would be effective theory 
 with a cutoff around the GUT scale, far below the Planck scale. 

In order to solve these problems, we have considered 
 the minimal SO(10) model in the warped extra dimension. 
As a simple setup, we have assumed that matter multiplets 
 reside on the IR brane 
 while the Higgs multiplets reside in the bulk. 
The warped geometry leads to a low scale effective cutoff
 in effective four dimensional theory, 
 and we fix it at the GUT scale. 
Therefore, the four dimensional minimal SO(10) model 
 is realized as the effective theory with the GUT scale cutoff.

\section*{Acknowledgments}
We are grateful to Y. Koide, the chairman of 
the International Workshop on Neutrino Masses and Mixings
Toward Unified Under standing of Quarks and Lepton Mass Matrices,
held at University of Shizuoka on December 17-19, 2006 for warm hospitality.
The work of T.F. and N.O. is supported in part by 
 the Grant-in-Aid for Scientific Research from the Ministry 
 of Education, Science and Culture of Japan (\#16540269, \#18740170).

\end{document}